\documentclass[%
 reprint,
 amsmath,assume,
 aps,
]{revtex4-2}
\usepackage{mathrsfs} 
\usepackage{graphicx}
\usepackage{dcolumn}
\usepackage{lipsum}
\usepackage{bm}
\usepackage[usenames]{color}
\usepackage{colortbl}
\usepackage{hyperref}
\usepackage{lineno}
\usepackage{soul}

\begin{document}

\title{
Reconstructing Critical Current Density in Josephson Junctions with Phase Non-linearity
}

\author{A. Kudriashov$^{1,2}$}
\email{andrei.kudriashov.97@gmail.com}
\author{R. A. Hovhannisyan$^{3}$}
\author{X. Zhou$^{1,2}$}
\author{L. Elesin$^{1,2,4}$}
\author{L. V. Yashina$^{5,6}$}
\author{K. S. Novoselov$^{1}$}
\author{D. A. Bandurin$^{1,2}$}
\email{dab@nus.edu.sg}

\affiliation{$^1$Institute for Functional Intelligent Materials, National University of Singapore, Singapore 117575, Singapore}
\affiliation{$^2$Department of Materials Science and Engineering, National University of Singapore, Singapore 117575, Singapore}
\affiliation{$^3$Department of Physics, Stockholm University, AlbaNova University Center, SE-10691 Stockholm, Sweden}
\affiliation{$^4$Programmable Functional Materials Lab, Center for Neurophysics and Neuromorphic Technologies, Moscow 127495}
\affiliation{$^5$Chemistry Department, M.V. Lomonosov Moscow State University, Moscow, Russia}
\affiliation{$^6$Moscow Center for Advanced Studies, Moscow, Russia.}

\begin{abstract}
In this Letter, we show that the standard Dynes–Fulton analysis, commonly used to reconstruct the critical current density from interference patterns, breaks down in Josephson junctions with nonlinear phase distributions, leading to non-physical artifacts.
To address this, we developed a simple iterative reconstruction algorithm and validated it both numerically and experimentally using a planar Josephson junction model.
Unlike conventional approaches based on the logarithmic
Hilbert transform, the proposed method allows for incorporating prior knowledge about the system and addresses the fundamental issue of ambiguity in reconstructing the critical current density from interference patterns.
\end{abstract}

\maketitle

Josephson junctions (JJs) are a versatile platform for probing a wide range of quantum phenomena, from unconventional superconductivity to topological states of matter. 
In particular, the spatial distribution of the critical current density across a junction can carry signatures of exotic physics, such as edge-dominated transport~\cite{bocquillon2017gapless,hart2014induced,diez2023symmetry} or unconventional pairing~\cite{fermin2022superconducting}. 
One of the most accessible experimental probes of this distribution is the reconstruction of the critical current density from the magnetic interference pattern, which is obtained by measuring the critical current as a function of an applied magnetic field.

Nowadays, a variety of approaches have been developed to achieve such a reconstruction~\cite{barone1982physics,dynes1971supercurrent,zappe1975determination,hui2014proximity,hart2014induced}, typically assuming a linear phase distribution across the junction length. While this assumption holds in many cases, certain experimental conditions naturally lead to nonlinear phase behavior. Such nonlinearities can arise from geometrical effects such as asymmetric inline junctions~\cite{guarcello2024efficiency}, inhomogeneities in the weak link~\cite{krasnov1997fluxon}, spatial inhomogeneities in the externally applied magnetic field~\cite{hovhannisyan2023superresolution}, or non-local electrodynamics~\cite{likharev2022dynamics,clem2010josephson,ivanchenko1990nonlocal}.

An extension of the Dynes–Fulton analysis was recently proposed in Ref.~\cite{fermin2023beyond}, where the phase distribution is linearized via a spatial rescaling transformation. While this method offers a way to address phase nonlinearity, it—like other approaches—suffers from the issue of non-uniqueness. This is a well-known problem in optics\cite{zappe1975determination,huang2016phase,hayes1982reconstruction,jaganathan2016phase,dainty1984essential}, and a variety of approaches have been developed to overcome it, such as the iterative Gerchberg–Saxton~\cite{gerchberg1972practical} and the Fienup~\cite{fienup1982phase} algorithms, both of which allow for the incorporation of prior knowledge about the object during reconstruction. However, unlike many optical tasks where such prior knowledge may be inaccessible, the Josephson junction is a unique system whose symmetries and constraints can often be inferred directly from its geometry or other characteristics. However, to our knowledge, such a method hasn't yet been developed.

In this Letter, we present an alternative method for reconstructing the critical current density distribution based on a straightforward iterative phase retrieval algorithm. 
First, using the example of a planar Josephson junction, we demonstrate that the standard reconstruction algorithm produces non-physical artifacts, which can be misinterpreted as edge-states.
Second, we solve the inverse problem for an arbitrary phase distribution along the junction length under an applied magnetic field, and develop a simple iterative algorithm to reconstruct the critical current density. 
Third, we show that incorporating additional prior knowledge about the junction properties addresses the issue of non-uniqueness in the inverse problem.
Finally, we validate the applicability of our method to an actual device - a planar Josephson junction based on an NbSe$_2$/Bi$_2$Se$_3$/NbSe$_2$ heterostructure.

\begin{figure*}
    \centering
    \includegraphics[width=0.9\textwidth]{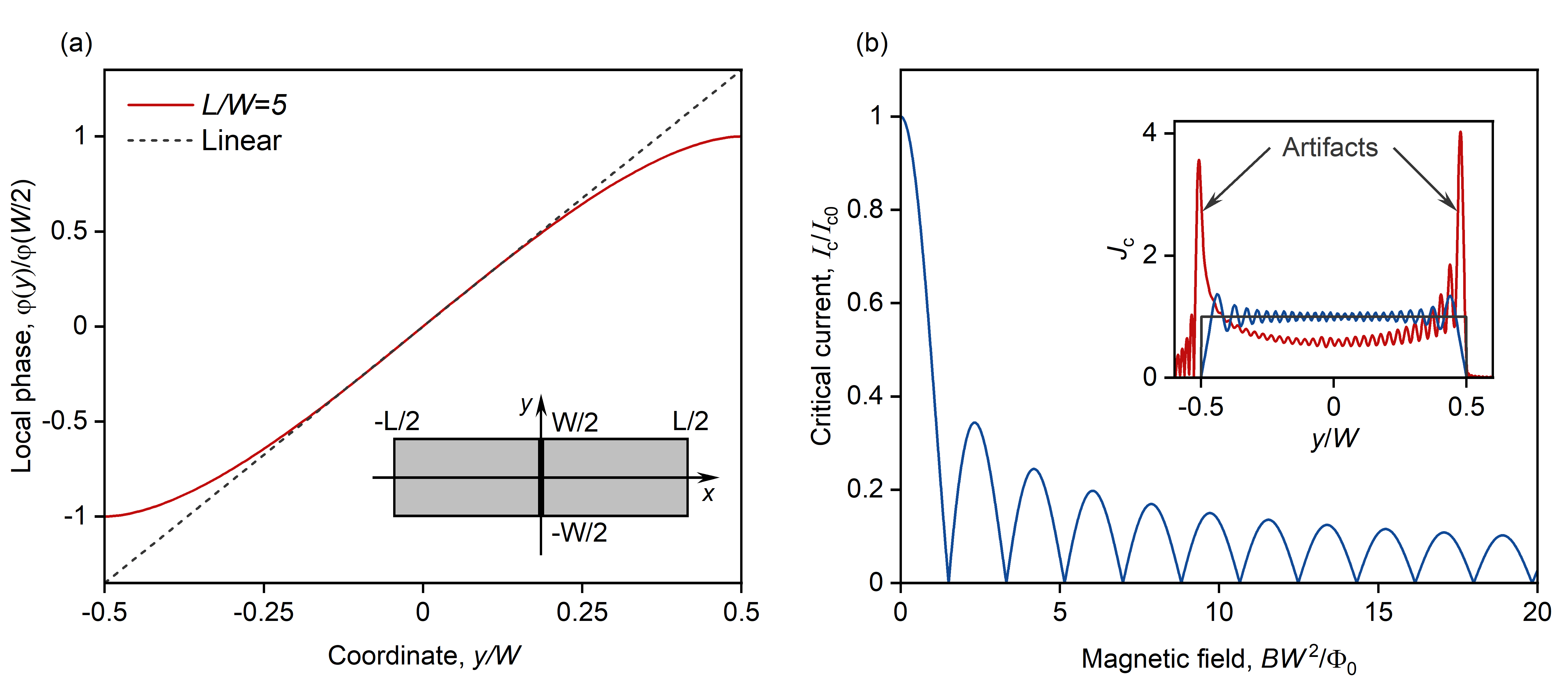}
    \caption{
    (a) Local phase difference between superconducting electrodes as a function of position along the planar Josephson junction, calculated using Eq.\eqref{phase}. 
    The dashed line shows a linear function for comparison.
    The insert illustrates the device geometry: gray regions represent superconducting electrodes of width $W$ and length $L/2$, separated by a weak link in the middle.
    (b) Magnetic interference pattern of the planar Josephson junction with $L/W=5$, calculated using Eq.\eqref{Ic}.  
    The insert shows the critical current density distribution: 
    the uniform $J_c(y)$ that was used in the simulation (black line),
    $J_c(y)$ obtained by the standard reconstruction algorithm (red line),
    $J_c(y)$ obtained by the reconstruction algorithm developed in this work (blue line).}
    \label{Fig.1}
\end{figure*}

As an example of a non-linear phase distribution along the junction length, we consider a planar Josephson junction with sinusoidal current-phase relation, consisting of two superconducting electrodes with length $L/2$ and width $W$, separated by a weak link, as shown in the insert of Fig.~\ref{Fig.1}(a).
The critical current $I_c$ of such a junction can be expressed as a maximum value of the integrated local supercurrent:
\begin{equation}\label{Ic}
I_c(B)
= \underset{\text{$\phi$}}{\text{max}} 
\int\limits_{-W/2}^{W/2}
J_c(y)
\sin(\varphi(y)\cdot B + \phi)
\mathrm{d}y
,
\end{equation}
where $y$ is a coordinate along the junction, $\varphi(y)\cdot B$ is a local phase difference between superconducting electrodes at a given magnetic field $B$, $\phi$ is the Josephson free phase, and $J_c(y)$ is the spatial distribution of the critical current density across a junction.

In considered planar geometry, an external magnetic field induces screening currents in the superconducting electrodes\cite{clem2010josephson}, and the resulting local phase difference $\varphi(y)$ can be expressed as the Fourier series:
\begin{equation} \label{phase}
    \varphi(y) =  \frac{16\pi}{\Phi_0 W}\sum_{n=0}^{\infty} \frac{(-1)^n}{k_n^3}\text{tanh}(k_n L/4) \text{sin}(k_n y),
\end{equation}
where $k_n=\frac{2\pi}{W}(n+1/2)$, $n$ is a non-negative integer, and $\Phi_0$ is the superconducting magnetic flux quantum.
Depending on the value of $L/W$, the local phase difference $\varphi(y)$ transitions from linear behavior for short and wide electrodes ($L/W \ll 1$) to highly non-linear behavior for long electrodes\cite{clem2010josephson}.
To illustrate this non-linearity, we plot $\varphi(y)$ for $L/W=5$ in Fig.\ref{Fig.1}(a).

\begin{figure*}[ht!]
    \centering
    \includegraphics[width=\textwidth]{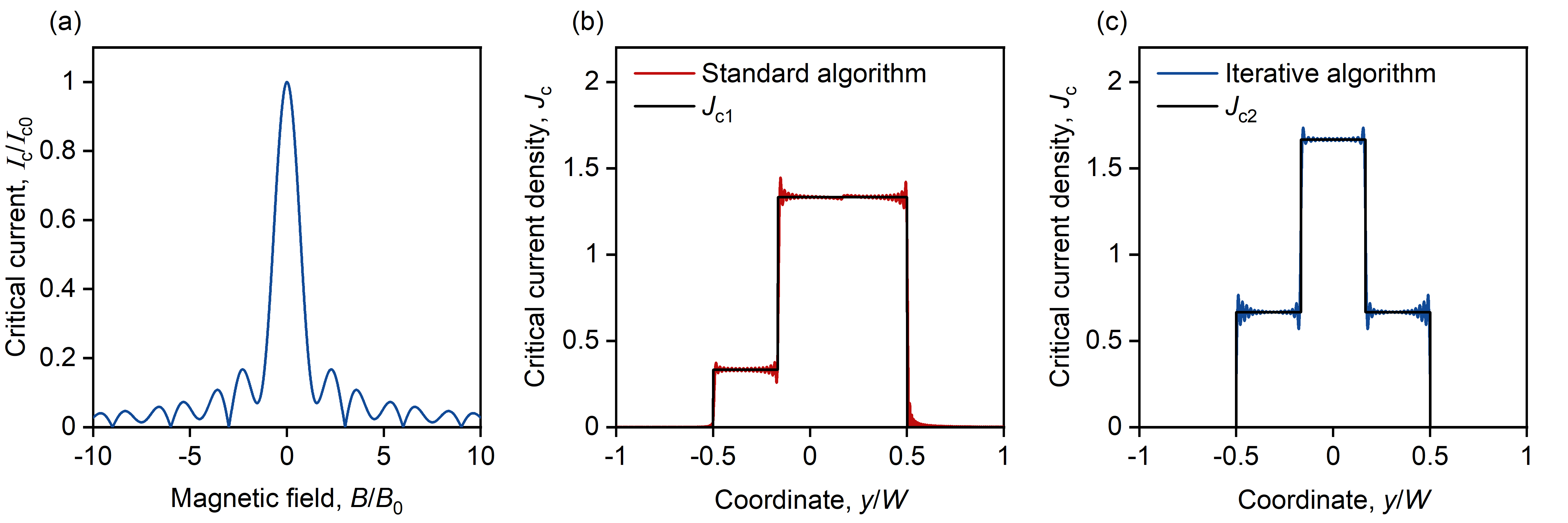}
    \caption{(a) Simulated magnetic interference pattern of the Josephson junction, which corresponds to both $J_{c1}(y)$ and $J_{c2}(y)$, shown by the black line in (b) and (c), respectively.
    (b) Critical current density distribution, obtained by the standard reconstruction algorithm (red line). 
    (c) Critical current density distribution, obtained by the iterative algorithm, enforcing it to be symmetric (blue line).} 
    \label{Fig:2}
\end{figure*}

The standard reconstruction algorithm relies on the assumption that local phase difference is a linear function of coordinate $\varphi(y,B)=\frac{2\pi}{\Phi_0}B y d_{eff}$, where $d_{eff}$ is the effective magnetic thickness of the device~\cite{dynes1971supercurrent}.
Therefore, this reconstruction algorithm can be safely applied to the planar Josephson junctions with very short and wide electrodes.
However, when this assumption is violated, it can lead to artifacts in the extracted critical current density $J_c(y)$. 
To illustrate this, we calculate the magnetic interference pattern $I_c(B)$ for uniform critical current density distribution $J_c(y)=1$, using Eqs.~\eqref{Ic} and \eqref{phase} for $L/W=5$.
The resulting $I_c(B)$ is shown in Fig.~\ref{Fig.1}(b).
Then, we apply the standard reconstruction algorithm and plot the resulting $J_c(y)$ in the insert of Fig.~\ref{Fig.1}(b) by the red line.
It deviates significantly from the uniform  $J_c(y)$, particularly near the edges of the Josephson junction, where an artificial enhancement of the critical current density is observed. 
Indeed, in the limit \( L/W \rightarrow \infty \), the \( I_c(B) \) pattern approaches the behavior of the zeroth-order Bessel function, whose inverse Fourier transform takes the form \( \frac{2}{\pi\sqrt{1 - 4y^2}} \) for \( |y| < \frac{1}{2} \). 
This function exhibits singularities at the edges, which can be misinterpreted as edge states~\cite{kudriashov2025non}. 

The correct solution can be obtained by first rewriting Eq.~(\ref{Ic}) in the form of a Fourier-like integral transform~\cite{hovhannisyan2023superresolution,dynes1971supercurrent,fermin2023beyond, supplemental}:
\begin{equation}\label{Direct}
iI_c(B) = \int\limits_{-W/2}^{W/2} J_c(y) e^{i B \varphi(y) + i\phi^*(B)}\,\mathrm{d}y,
\end{equation}
where $\phi^*(B)$ is the critical phase, i.e., the phase at which the critical current is achieved.

Then, the solution to the inverse problem—reconstructing $J_c(y)$ from $I_c(B)$—can be expressed as:
\begin{equation}\label{Inverse}
J_c(y) = \frac{|\varphi'(y)|}{2\pi} \int\limits_{-\infty}^{\infty} iI_c(B) e^{-i B \varphi(y) - i\phi^*(B)}\,\mathrm{d}B,
\end{equation}
where $\varphi'(y) = \frac{d\varphi(y)}{dy}$ (see Supplementary Materials for details\cite{supplemental}).


To achieve the full reconstruction using Eq.~(\ref{Inverse}), one needs to know the critical phase $\phi^*$. Inspired by the Gerchberg–Saxton~\cite{gerchberg1972practical} algorithm, we have developed a simple iterative scheme for phase reconstruction and determination of the critical current density consisting of the following steps:
 
1) \textit{Initial guess for critical phase:} 
a good starting point is the phase profile for a uniform critical current density $J_c(y)=1$, where $\phi^* = \pm \pi/2$, depending on the lobe of the critical current oscillation.

2) \textit{Reconstruction of complex current density:} 
using $\phi^*(B)$, apply Eq.~\eqref{Inverse} to reconstruct an approximate (generally complex) critical current density profile $J_c(y)$.

3) \textit{Enforce physical constraints:}  
the actual critical current density must be real-valued and positive: $J_c^{new}(y)=|Re(J_c(y))|$.

4) \textit{Update critical phase:}
with the improved estimate $J_c^{new}(y)$, recalculate the critical phase $\phi^*(B)$ using Eq.\eqref{Direct}.

5) \textit{Iterate until convergence:}
repeat steps 2–4, updating and refining both $\phi^*(B)$ and $J_c(y)$, until the procedure converges to a stable solution.

To demonstrate the validity of this algorithm, we apply it to reconstruct the supercurrent density distribution from the calculated $I_c(B)$, shown in Fig.~\ref{Fig.1}(b). 
The resulting $J_c(y)$ is shown in the insert of Fig.~\ref{Fig.1}(b) by the blue line. 
Compared to the standard algorithm (red line in the insert of Fig.~\ref{Fig.1}(b)), the iterative algorithm correctly reconstructs the critical current density distribution without introducing artifacts. During testing, we applied the reconstruction procedure to a variety of $J_c(y)$ profiles, confirming the robustness and correctness of the method (see Supplementary Materials for more\cite{supplemental}).
We note that Eq.~(\ref{phase}) is valid for rectangular electrodes, but the reconstruction algorithm is general and can be applied to any known phase profile $\varphi(y)$, which can be computed numerically for arbitrary electrode geometries using Ginzburg–Landau theory\cite{clem2010josephson,fermin2023beyond}.

However, the constraints applied in the third step of the iteration process are not sufficient to resolve the fundamental issue of non-uniqueness inherent in inverse problems~\cite{huang2016phase,hayes1982reconstruction,jaganathan2016phase}. As demonstrated by Zappe~\cite{zappe1975determination}, the critical current versus magnetic field dependence \( I_c(B) \), shown in Fig.~\ref{Fig:2}(a), can correspond to two distinct real positive profiles of the critical current density \( J_c(y) \), as illustrated by the black lines in Fig.~\ref{Fig:2}(b,c), thereby making a fully unique reconstruction impossible without prior knowledge about $J_c$.

The logarithmic Hilbert transform, used as the primary phase retrieval technique, assumes that the object under consideration is of minimum phase~\cite{oppenheim1983signals}. As a result, it reconstructs only the profile of the critical current density shown in Fig.~\ref{Fig:2}(b) and thus may lead to an incorrect solution.

A commonly used approach to address the non-uniqueness of the inverse problem involves assuming high symmetry in the critical current density profile \( J_c(y) \). In several previous studies~\cite{hart2014induced,hui2014proximity,fermin2022superconducting}, the reconstruction was performed in two steps: first, \( J_c(y) \) was obtained under the assumption of oddness, and then a small asymmetric component was added to improve agreement with experimental data. While this method can produce reasonable approximations, it depends on the specific form of the assumed asymmetry and does not fully resolve the ambiguity of the inverse problem~\cite{hui2014proximity}.

\begin{figure*}[ht!]
    \centering
    \includegraphics[width=\textwidth]{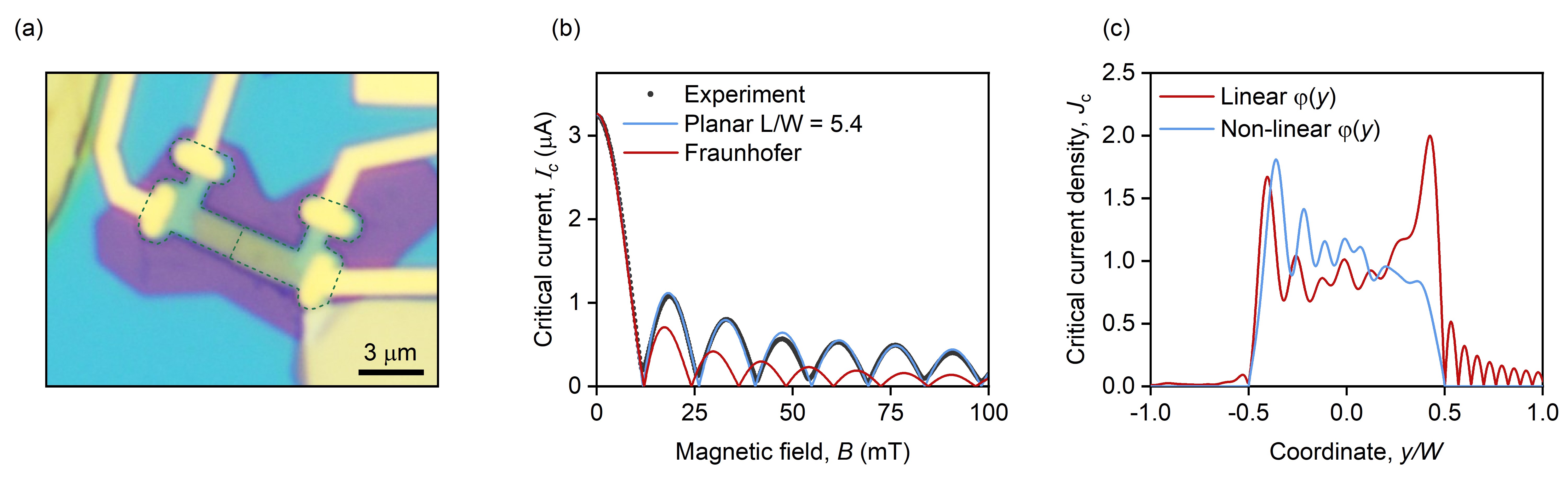}
    \caption{(a) Optical microscopy image of a planar Josephson junction fabricated from a cracked NbSe$_2$ flake as the superconducting electrodes (green dashed lines) and a few-quintuple-layer Bi$_2$Se$_3$ flake as the weak link.  
(b) Experimentally measured dependence of the critical current $I_c(B)$ on the applied magnetic field (black), compared with numerically calculated $I_c(B)$ using rectangular model with a nonlinear phase distribution, given by eq.~(\ref{phase}), for aspect ratio $L/W \simeq 5.4$ (blue), and with linear phase distribution and uniform critical current density (red).
(c) Reconstructed critical current density distribution along the junction using the standard (red line) and developed in the work (blue line) methods.} 
    \label{Fig:3}
\end{figure*}

In contrast, the iterative approach developed in this work, based on the iterative algorithms, offers a more flexible framework for incorporating prior knowledge. 
For example, if it is known that the junction should exhibit a mostly symmetric \( J_c(y) \), this constraint can be directly enforced during the iterative process. 
This allows the algorithm to converge toward a physically meaningful and self-consistent solution. 
Figure~\ref{Fig:2}(c) presents the result of such a calculation for the example presented by Zappe, where the iterative method successfully reconstructs a symmetric profile consistent with the known properties of the junction. 
Note that the algorithm developed in this work is capable of reconstructing \( J_{c1} \) (see Supplementary Materials for details\cite{supplemental}), thus helping to resolve the problem of non-uniqueness.

To validate the applicability of our method to an actual device, we fabricated a planar superconductor–normal metal–superconductor (SNS) Josephson junction using an NbSe$_2$/Bi$_2$Se$_3$/NbSe$_2$ heterostructure (see Fig.~\ref{Fig:3}(a)). 
Fabrication details can be found in the Ref.~\cite{kudriashov2025non}.
We measure $I_c(B)$ dependence of this device at $T=10$~mK and show it in Fig.~\ref{Fig:3}(b) by black dots. 
Then, we calculate the theoretical $I_c(B)$ assuming uniform critical current distribution for two models: one assuming a linear phase gradient (shown by a red line), and one using the Clem model as described by Eq.~(\ref{phase}) with $W = 1.625~\mu\mathrm{m}$ and $L = 8.7~\mu\mathrm{m}$ (shown by a blue line). 
While the Fraunhofer-like prediction based on a linear phase clearly fails to capture the experimental behavior (including periodicity of the oscillations), the Clem model provides a much closer fit, supporting its validity for real devices.

Next, we reconstruct the critical current density distribution using the standard linear-phase method (red line in Fig.~\ref{Fig:3}(c)) and our algorithm (blue line in Fig.~\ref{Fig:3}(c)). As expected, the linear method introduces spurious edge features, consistent with theoretical calculations (see the insert of Fig.\ref{Fig.1}(a)). On the contrary, our approach yields a more physically plausible distribution: nearly uniform $J_c(y)$ with a slight tilt, which can be attributed to the non-uniform distance between superconducting electrodes.
This result demonstrates the applicability of our approach to realistic device geometries and experimental data.


To conclude, we introduced a method for reconstructing the critical current density distribution using a straightforward phase retrieval algorithm. 
The approach involves solving the inverse problem for arbitrary phase distributions under an applied magnetic field, followed by a simple iterative algorithm to reconstruct the critical current density. 
Numerical and experimental validations for planar Josephson junctions confirm that incorrect reconstruction techniques can produce misleading artifacts, highlighting the importance of applying a correct reconstruction method.
To facilitate broader use of our technique, we provide open-source implementations of the algorithm in \texttt{MATLAB}, \texttt{Python}, and \texttt{LabVIEW}, making it accessible to the experimental superconducting electronics and quantum transport communities\cite{Scripts}.

\begin{acknowledgments}
This work was supported by the Singapore Ministry of Education AcRF Tier 2 grant T2EP50123-0020. K.S.N. and A.K. are grateful to the Ministry of Education, Singapore (Research Centre of Excellence award to the Institute for Functional Intelligent Materials, I-FIM, project No. EDUNC-33-18-279-V12) and to the Royal Society (UK, grant number RSRP$\setminus$ R$\setminus$ 190000) for support.
L.E. acknowledges the support of the internal funding programme from the Center for Neurophysics and Neuromorphic Technologies. 
\end{acknowledgments}

A.K. and R.A.H. contributed equally to this work.
R.A.H. and A.K. performed a theoretical analysis. L.V.Y. provided the Bi$_2$Se$_3$ crystals. X.Z., A.K, and L.E. fabricated the sample. A.K. performed transport measurements. 
A.K. and R.A.H wrote the manuscript with the contribution from all authors.
D.A.B. supervised the project.

\bibliography{Bibliography.bib}

\end{document}


\setcounter{figure}{0}
\renewcommand{\thesection}{}
\renewcommand{\thesubsection}{S\arabic{subsection}}
\renewcommand{\theequation} {S\arabic{equation}}
\renewcommand{\thefigure} {S\arabic{figure}}
\renewcommand{\thetable} {S\arabic{table}}

\title{Supplementary Materials for \\  Reconstructing Critical Current Density in Josephson Junctions with Phase Non-linearity}

\author{A. Kudriashov}
\affiliation{Institute for Functional Intelligent Materials, National University of Singapore, Singapore 117575, Singapore}
\affiliation{Department of Materials Science and Engineering, National University of Singapore, Singapore 117575, Singapore}

\author{R. A. Hovhannisyan}
\affiliation{Department of Physics, Stockholm University, AlbaNova University Center, SE-10691 Stockholm, Sweden}

\author{X. Zhou}
\affiliation{Institute for Functional Intelligent Materials, National University of Singapore, Singapore 117575, Singapore}
\affiliation{Department of Materials Science and Engineering, National University of Singapore, Singapore 117575, Singapore}

\author{L. Elesin}
\affiliation{Institute for Functional Intelligent Materials, National University of Singapore, Singapore 117575, Singapore}
\affiliation{Department of Materials Science and Engineering, National University of Singapore, Singapore 117575, Singapore}
\affiliation{Programmable Functional Materials Lab, Center for Neurophysics and Neuromorphic Technologies, Moscow 127495}

\author{L. V. Yashina}
\affiliation{Chemistry Department, M.V. Lomonosov Moscow State University, Moscow, Russia}
\affiliation{Moscow Center for Advanced Studies, Moscow, Russia.}

\author{K. S. Novoselov}
\affiliation{Institute for Functional Intelligent Materials, National University of Singapore, Singapore 117575, Singapore}

\author{D. A. Bandurin}
\affiliation{Institute for Functional Intelligent Materials, National University of Singapore, Singapore 117575, Singapore}
\affiliation{Department of Materials Science and Engineering, National University of Singapore, Singapore 117575, Singapore}

\maketitle

\tableofcontents

\subsection{Derivation of the direct transformation}
The supercurrent $I_s$ through the Josephson junction can be expressed as:
\begin{equation}\label{supercurrent}
I_s(B, \phi)
= 
\int\limits_{-W/2}^{W/2}
J_c(y)
\sin(\varphi(y)\cdot B + \phi)
\mathrm{d}y
,
\end{equation}
where $B$ is the external magnetic field, $\phi$ is a Josephson free phase, $W$ is a junction width, $y$ is a coordinate along the junction, and $J_c(y)$ is a critical current density distribution, and $\varphi(y)\cdot B$ is a local phase difference between the superconducting electrodes at a given magnetic field $B$.

The critical current $I_c$ is the maximum supercurrent at a given magnetic field $B$, therefore derivative of the supercurrent over the Josephson free phase must be equal to zero: 

\begin{equation}\label{derivative}
\left. \frac{\partial I_s(B, \phi)}{\partial \phi} \right|_{\phi=\phi^*}
=
\int\limits_{-W/2}^{W/2}
J_c(y)
\cos(\varphi(y)\cdot B + \phi^*)
\, \mathrm{d}y=0,
\end{equation}

\begin{equation}\label{critical_current}
I_c(B)= 
\int\limits_{-W/2}^{W/2}
J_c(y)
\sin(\varphi(y)\cdot B + \phi^*)
\mathrm{d}y,
\end{equation}
where $\phi^*$ is a critical phase, i.e., the phase at which the maximal supercurrent is achieved. 

Summing up eq.~\ref{derivative} and eq.~\ref{critical_current}, multiplied by the imaginary unit and using Euler's formula, we get:

\begin{equation}\label{sum}
0+iI_c(B) = \int\limits_{-W/2}^{W/2} J_c(y)[cos(\varphi(y)\cdot B+\phi^*(B))+i sin(\varphi(y)\cdot B+\phi^*(B))]\mathrm{d}y.
\end{equation}

\begin{equation}
iI_c(B) = \int\limits_{-W/2}^{W/2} J_c(y)e^{i B\varphi(y)+i\phi^*(B)}\mathrm{d}y.
\end{equation}

We can rewrite it as:

\begin{equation}
e^{-i \phi^*(B)} = \frac{\int\limits_{-W/2}^{W/2} J_c(y)e^{i B \varphi(y)}\mathrm{d}y}{iI_c(B)}.
\end{equation}

Since $I_c(B)$ is a real function, it does not change the value of the phase $\phi^*(B)$, therefore the final expression for $\phi^*(B)$ is:

\begin{equation}
\phi^*(B) = -\text{arg}(-i\int\limits_{-W/2}^{W/2} J_c(y)e^{i B \varphi(y)}\mathrm{d}y)
\end{equation}

\subsection{Proof that Equation~\ref{Inverse} inverts Equation~\ref{Direct}}
The inverse transformation is:
\begin{equation}\label{Inverse}
J_c(y) = \frac{|\varphi'(y)|}{2\pi} \int_{-\infty}^{\infty} i I_c(B) \, e^{-i B \varphi(y) - i \phi^*(B)} \, \mathrm{d}B \tag{4}
\end{equation}

The direct transformation is:
\begin{equation}\label{Direct}
i I_c(B) = \int_{-W/2}^{W/2} J_c(y') \, e^{i B \varphi(y') + i \phi^*(B)} \, \mathrm{d}y' \tag{3}
\end{equation}

Let us substitute \ref{Direct} into \ref{Inverse}:

\begin{equation}
J_c(y) = \frac{|\varphi'(y)|}{2\pi} \int_{-\infty}^{\infty} \left( \int_{-W/2}^{W/2} J_c(y') \, e^{i B \varphi(y') + i \phi^*(B)} \, \mathrm{d}y' \right) e^{-i B \varphi(y) - i \phi^*(B)} \, \mathrm{d}B
\end{equation}

\begin{equation}
 J_c(y) = \frac{|\varphi'(y)|}{2\pi} \int_{-W/2}^{W/2} J_c(y') \left( \int_{-\infty}^{\infty} e^{i B (\varphi(y') - \varphi(y))} \, \mathrm{d}B \right) \mathrm{d}y'
\end{equation}

The critical phase $\phi^*(B)$ cancels out. The inner integral is a Fourier representation of the Dirac delta function:

\begin{equation}
\int_{-\infty}^{\infty} e^{i B (\varphi(y') - \varphi(y))} \, \mathrm{d}B = 2\pi \delta(\varphi(y') - \varphi(y)) 
\end{equation}

Thus:
\begin{equation}
J_c(y) = |\varphi'(y)| \int_{-W/2}^{W/2} J_c(y') \, \delta(\varphi(y') - \varphi(y)) \, \mathrm{d}y' 
\end{equation}

Now, we apply the change-of-variable property of the Dirac delta function:
\begin{equation}
\delta(\varphi(y') - \varphi(y)) = \frac{\delta(y' - y)}{|\varphi'(y)|}  
\end{equation}
where $\varphi(y)$ assumed to be inseverable function. 

Substitute back:
\begin{equation}
J_c(y) = |\varphi'(y)| \int_{-W/2}^{W/2} J_c(y') \cdot \frac{\delta(y' - y)}{|\varphi'(y)|} \, \mathrm{d}y' = \int_{-W/2}^{W/2} J_c(y') \delta(y' - y) \, \mathrm{d}y' = J_c(y)  
\end{equation}

\textbf{Q.E.D.}

\subsection{Additional example of reconstruction}

\begin{figure*}[ht!]
    \centering
    \includegraphics[width=\textwidth]{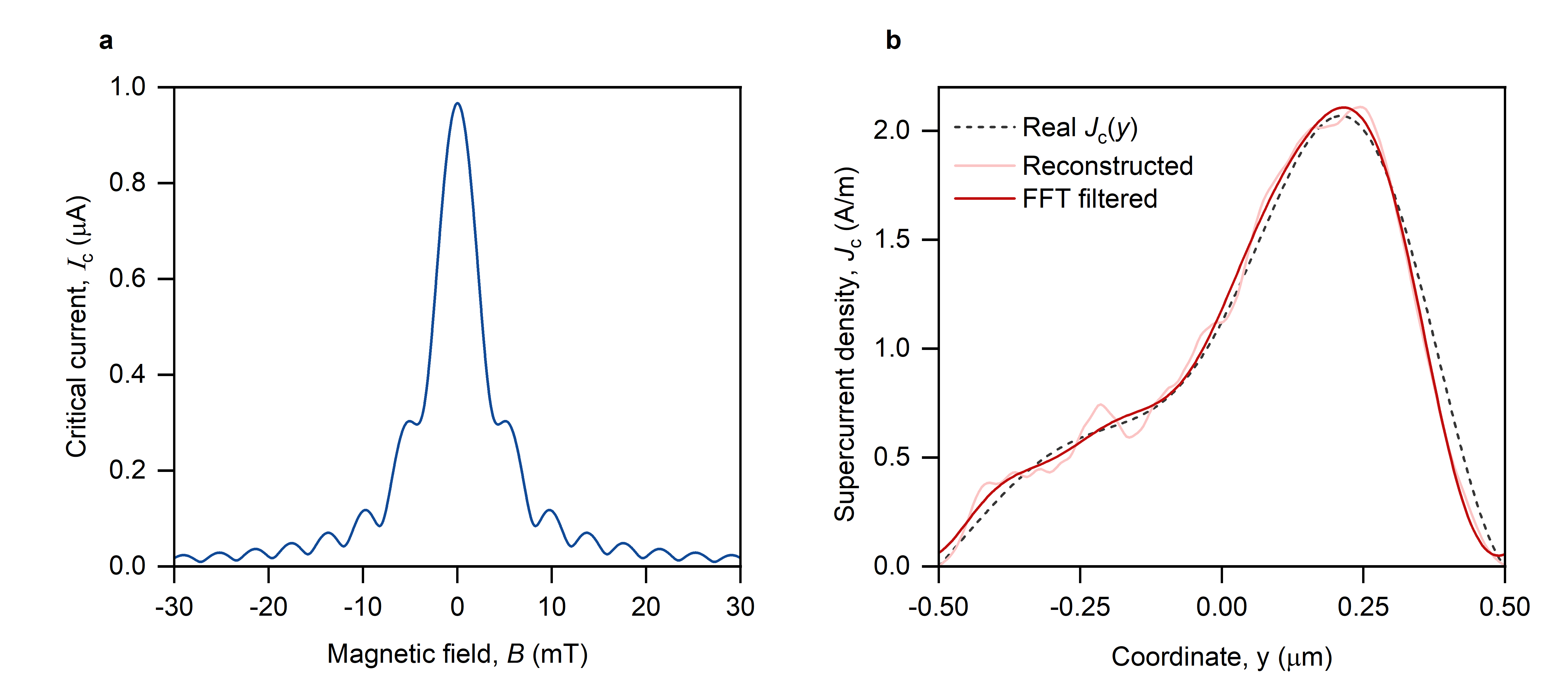}
    \caption{(a) Simulated magnetic interference pattern $I_c(B)$ of the planar Josephson junction with $L/W=5$ and $J_c(y)$, shown in (b) by the black dashed line.
    (b) Critical current density distribution, obtained by the iterative reconstruction algorithm developed in this work (pink line). Red line shows FFT-filtered $J_c(y)$. 
    } 
    \label{rec}
\end{figure*}

Figure~\ref{rec} illustrates the reconstruction of an asymmetric critical current density profile (see Fig.~\ref{rec}(b), black dashed curve) calculated for a nonlinear phase distribution in a planar Josephson junction with an electrode aspect ratio of $L/W = 5$. As evident, the $I_c(B)$ dependence is significantly distorted compared to the conventional Fraunhofer pattern, exhibiting notably elevated side lobes at the first minima.

In Fig.~\ref{rec}(b), the reconstructed current density is shown in light red curve. While the reconstructed profile shows good agreement with the original distribution, small inaccuracies appear in the form of oscillatory artifacts, similar to those observed in the main text. As discussed in Ref.~[11], these oscillations originate from the finite range of the $I_c(B)$ dependency, which limits the precision of the reconstruction.

However, such artifacts can be effectively suppressed by applying low-pass Fourier filtering techniques. The thick red curve in the same figure shows the resulting $J_c(y)$ after applying image enhancement methods. As observed, the filtered reconstruction provides a much more accurate and visually improved representation of the original current density.

\subsection{Convergence of iterative algorithm}
\begin{figure*}[ht!]
    \centering
    \includegraphics[width=\textwidth]{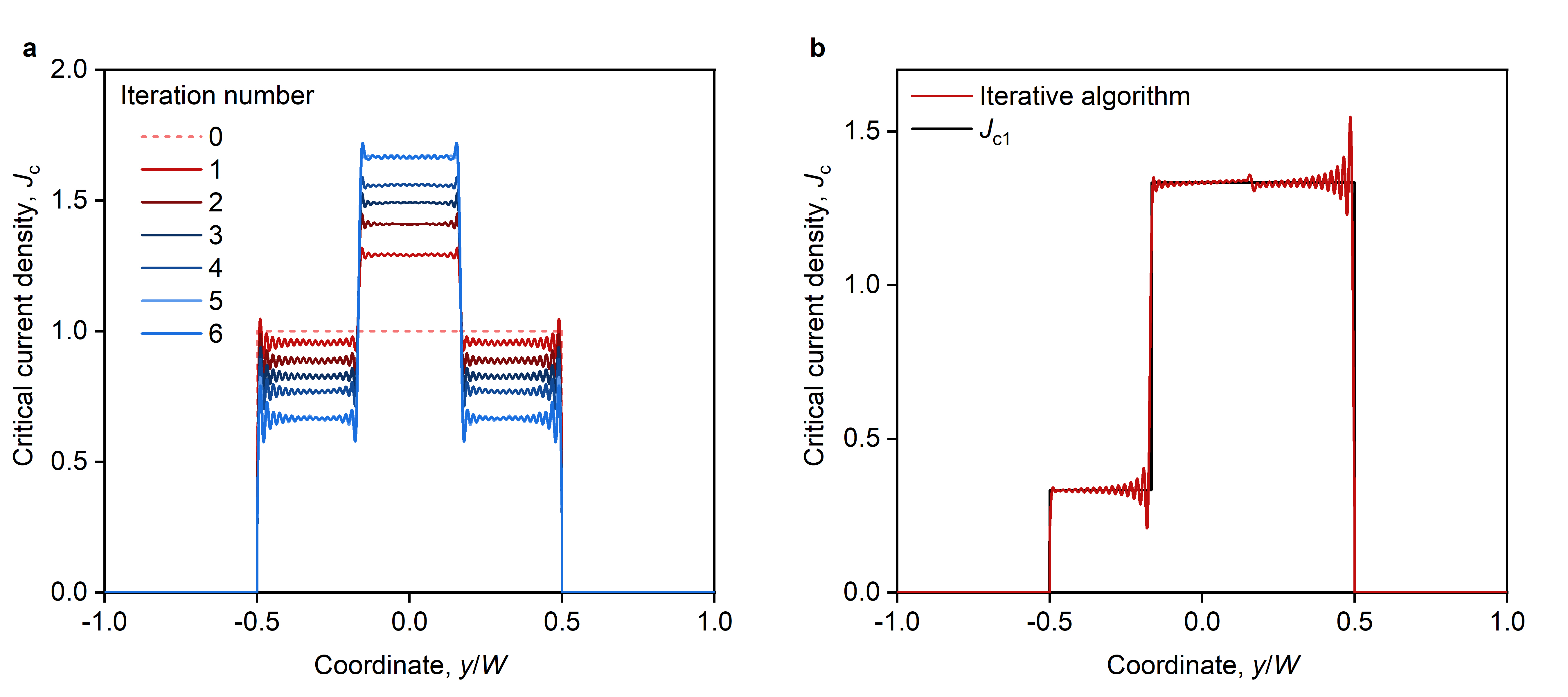}
    \caption{(a) Reconstructed critical current density at various stages of algorithm convergence. (b) Critical current density $J_{c1}$, reconstructed by iterative algorithm. 
    } 
    \label{SupFig2}
\end{figure*}

Fig.~\ref{SupFig2}(a) shows the profiles of $J_c(y)$ calculated for the dependence of $J_{c2}$ (see Fig.~2(c) of the main text) at different steps of the iterative process. The initial guess for $J_c(y)$ was taken to be uniform, and symmetry constraints were applied to the overall form of the current density distribution in the form
\[
J_{sym}^{i}(y) = \frac{1}{2}\left(J_c^{\text{rec}}(y) + J_c^{\text{rec}}(-y)\right),
\]
where $J_c^{\text{rec}}$ denotes the reconstructed solution obtained from the inverse problem. One can observe that in the initial iterations, the reconstructed $J_{sym}^i(y)$ significantly deviates from the true profile, although it captures the overall qualitative behavior. After only a few iterations (six in total), the curve converges to the true profile of $J_c(y)$.

Figure~\ref{SupFig2}(b) shows the iteratively reconstructed profile of $J_{c1}$ as a red curve. To achieve this reconstruction, we incorporated the knowledge that the overall solution should have an asymmetric component, using an initial guess for $J_c(y)$ in the form of a linear tilt: $J_c(y) = 1 - y$. In this case, accurate reconstruction and convergence were achieved by the 20th iteration step.

\newpage
\setcounter{figure}{0}
\renewcommand{\thesection}{}
\renewcommand{\thesubsection}{S\arabic{subsection}}
\renewcommand{\theequation} {S\arabic{equation}}
\renewcommand{\thefigure} {S\arabic{figure}}
\renewcommand{\thetable} {S\arabic{table}}